\documentclass{article}

\usepackage{arxiv}

\usepackage[utf8]{inputenc} 
\usepackage[T1]{fontenc}    
\usepackage{hyperref}       
\usepackage{url}            
\usepackage{booktabs}       
\usepackage{amsfonts}       
\usepackage{nicefrac}       
\usepackage{microtype}      
\usepackage{lipsum}

\usepackage{blindtext}

\usepackage[maxfloats=100]{morefloats}
\usepackage{relsize}

\usepackage{amssymb}

\usepackage{comment}
\usepackage{lineno}
\newcommand*\patchAmsMathEnvironmentForLineno[1]{%
  \expandafter\let\csname old#1\expandafter\endcsname\csname #1\endcsname
  \expandafter\let\csname oldend#1\expandafter\endcsname\csname end#1\endcsname
  \renewenvironment{#1}%
     {\linenomath\csname old#1\endcsname}%
     {\csname oldend#1\endcsname\endlinenomath}}%
\newcommand*\patchBothAmsMathEnvironmentsForLineno[1]{%
  \patchAmsMathEnvironmentForLineno{#1}%
  \patchAmsMathEnvironmentForLineno{#1*}}%
\AtBeginDocument{%
\patchBothAmsMathEnvironmentsForLineno{equation}%
\patchBothAmsMathEnvironmentsForLineno{align}%
\patchBothAmsMathEnvironmentsForLineno{flalign}%
\patchBothAmsMathEnvironmentsForLineno{alignat}%
\patchBothAmsMathEnvironmentsForLineno{gather}%
\patchBothAmsMathEnvironmentsForLineno{multline}%
}

\usepackage{graphicx}
\graphicspath{{Figures_png/}}
\usepackage{float}
\usepackage{caption}
\usepackage{subcaption}
\usepackage{amsmath}
\usepackage{multirow}

\usepackage{color}
\definecolor{rev}{rgb}{0,0,0} 

\usepackage{graphics}
\usepackage{adjustbox}
\usepackage{algorithm}
\usepackage[noend]{algpseudocode}



\usepackage{hyperref}

\title{A nudged hybrid analysis and modeling approach for realtime wake-vortex transport and decay prediction}

\author{
 Shady E. Ahmed \\
  School of Mechanical \& Aerospace Engineering,\\
  Oklahoma State University,\\
  Stillwater, OK 74078, USA.\\
  \texttt{shady.ahmed@okstate.edu}
\And
 Suraj Pawar \\
  School of Mechanical \& Aerospace Engineering,\\
  Oklahoma State University,\\
  Stillwater, OK 74078, USA.\\
  \texttt{supawar@okstate.edu } 
\And
 Omer San \\
  School of Mechanical \& Aerospace Engineering,\\
  Oklahoma State University,\\
  Stillwater, OK 74078, USA.\\
  \texttt{osan@okstate.edu} 
\And
 Adil Rasheed \\
  Department of Engineering Cybernetics,\\
  Norwegian University of Science and Technology,\\
  N-7465, Trondheim, Norway.\\
  \texttt{adil.rasheed@ntnu.no } \\
 \And
 Mandar Tabib \\
 CSE Group, Mathematics and Cybernetics,\\
 SINTEF Digital,\\ 
 7034, Trondheim, Norway \\
 \texttt{mandar.tabib@sintef.no}
}

\begin{document}
\maketitle

\begin{abstract}
We put forth a long short-term memory (LSTM) nudging framework for the enhancement of reduced order models (ROMs) of fluid flows utilizing noisy measurements for air traffic improvements. Toward emerging applications of digital twins in aviation, the proposed approach allows for constructing a realtime predictive tool for wake-vortex transport and decay systems. We build on the fact that in realistic application, there are uncertainties in initial and boundary conditions, model parameters, as well as measurements. Moreover, conventional nonlinear ROMs based on Galerkin projection (GROMs) suffer from imperfection and solution instabilities, especially for advection-dominated flows with slow decay in the Kolmogorov $n$-width. In the presented LSTM nudging (LSTM-N) approach, we fuse forecasts from a combination of imperfect GROM and uncertain state estimates, with sparse Eulerian sensor measurements to provide more reliable predictions in a dynamical data assimilation framework. We illustrate our concept by solving the two-dimensional vorticity transport equation. We investigate the effects of measurements noise and state estimate uncertainty on the performance of the LSTM-N behavior. We also demonstrate that it can sufficiently handle different levels of temporal and spatial measurement sparsity, and offer a huge potential in developing next-generation digital twin technologies for aerospace applications.
\end{abstract}


\section{Introduction} \label{sec:intro}
Aircraft wings are optimized to produce maximum lift and minimum drag. Their design ensures that there is a high-pressure zone below the wing and a low-pressure zone above. Owing to this pressure gradient, air from below the wing is drawn around the wingtip into the region above the wing causing a vortex to trail from each wing tip. These wake vortices (WVs) are stable under calm atmospheric conditions and remain present in the free atmosphere for a very long time, retaining its shape and energy \cite{holzapfel2003analysis,holzapfel2003strategies,breitsamter2011wake}. Furthermore, at very low altitudes, they might rebound from the ground and linger on in the flight path corridor, posing significant risks to other aircraft that might encounter them. This is, in particular, crucial for large jetliners as WVs can cause violent rolling motions and even flip a small aircraft upside down when a pilot trailing a large aircraft flies into such vortices \cite{luckner2004hazard}.

It is due to the hazards posed by WVs left behind by a taking off or landing aircraft that serious precautions are to be taken. The operational minimum aircraft separation for different weight class configurations, used by the Air Traffic Control (ATC), varies from 2.5 to 6 nautical miles. However, when deciding the separation distance following those guidelines, the weather conditions and associated transport and decay of WVs are not often taken into account. This was not a serious issue a couple of decades ago, but with the significant increase of the air traffic and a push for remote towers for cost effective and safe operation, major airports around the world are feeling the pressure. In this regard, Digital Twins of airports appear like a potential solution. In the current context, there is a need to develop a more efficient wake turbulence separation consisting of time-based minima between different aircraft types which takes into account the dynamic meteorological factors along with the variation in the wake generation mechanism associated with different classes of aircraft. Such information will enable air traffic controllers to deliver consistent and safe spacing between aircraft leading to increased airport capacity, enhanced safety, reduced fuel consumption, improved predictability and increased resilience. 

While the current solutions range from actively modifying/dissipating the wake-vortices using physical devices \cite{anton2017noo,holzaepfel2020plr} to accurately estimating the strengths of the vortices using LIDARS and RADARS \cite{gerz2002commercial,holzapfel2003strategies}, one shortfall of the two approaches is that none of them predicts the evolution of the vortices in the future. This gap is being filled by advanced computational fluid dynamics modeling which involves solving the highly non-linear Navier-Stokes equations at varying levels of approximations. However, their utility owing to their computationally demanding nature has been limited to offline simulations geared towards developing a better understanding of the WV dynamics. At the moment, most of the fast WV models that are state-of-the art in WV predictive systems use physics-based empirical parameterizations to mimic vortex transport and decay. Unfortunately, the computational efficiency of the fast WV models comes at the expense of accuracy. A good overview of the models can be found in \cite{hallock2018aro}. To alleviate the problems associated with the existing WV models, data-driven machine learning methods might appear attractive at a first glance, but their limited interpretability owing to their black-box nature make them a misfit for the kind of safety-critical application under consideration. 

To this end, building upon our recent works on the hybrid analysis and modeling (HAM) framework \cite{pawar2020data,pawar2020evolve,ahmed2020long}, we present a data assimilation-empowered approach to utilize a machine learning methodology to fuse computationally-light physics-based models with the available real-time measurement data to provide more accurate and reliable predictions of wake-vortex transport. In particular, we build a surrogate reduced order model (ROM), by combining proper orthogonal decomposition (POD) for basis construction \cite{berkooz1993proper,chatterjee2000introduction,liang2002proper,rathinam2003new,kerschen2005method,volkwein2011model} and Galerkin projection to model the dynamical evolution on the corresponding low-order subspace \cite{rowley2004model,rapun2010reduced,lorenzi2016pod,kunisch2002galerkin,noack2011reduced,huang2016exploration,rezaian2020impact,kunisch2001galerkin,xu2020pod}. Although ROMs based on Galerkin projection (denoted as GROMs in the present study) have been traditionally considered the standard approach for reduced order modeling, they often become inaccurate and unstable for long-term predictions of convection-dominated flows with strong nonlinearity \cite{kalashnikova2010stability,grimberg2020stability,lassila2014model,rempfer2000low,noack2003hierarchy}. Ideas like closure modeling \cite{sirisup2004spectral,san2014proper,san2014basis,protas2015optimal,cordier2013identification,osth2014need,kalb2007intrinsic,xie2018data,mohebujjaman2019physically,akhtar2012new,balajewicz2012stabilization,amsallem2012stabilization,san2015stabilized,gunzburger2019evolve,wang2011two,iliescu2014variational,xie2017approximate,xie2018numerical,rahman2019dynamic,rezaian2020hybrid,rezaian2020multi,imtiaz2020nonlinear} and Petrov-Galerkin projection \cite{carlberg2011efficient,carlberg2017galerkin,wentland2019closure,choi2019space,lozovskiy2017evaluation,collins2020petrov,parish2020adjoint,xiao2013non,fang2013non} have been investigated to address this deficiency. Alternatively, we exploit the nudging method \cite{lakshmivarahan2013nudging} as a data assimilation (DA) framework, which works by relaxing the model state toward observations by adding correction (or nudging) terms, proportional to the difference between observations and model state, known as innovation in DA context. In classical DA nudging, this proportionality is assumed to be linear, and the proportionality constants (or weights) are empirically tuned. Instead, we introduce the hybridization at this stage, using a simplistic long short-term memory (LSTM) architecture to generalize this relation to consider nonlinear mappings among the innovation and nudging terms. 

In other words, we utilize LSTM to combine the possibly defective model prediction with noisy measurements to ``nudge''  the model's solution towards the true states \cite{pawar2020long,ahmed2020reduced}. We apply the proposed LSTM nudging framework (denoted LSTM-N) for the reduced order modeling of the two-dimensional wake vortex problem in order to accurately predict the transport of wake vortices. Moreover, we suppose that both inputs (i.e., the physics-based model and data) are imperfect, thus avoiding biases in predictions. GROMs are inherently imperfect due to the modal truncation and intrinsic nonlinearity. \textcolor{rev}{Recurrent neural networks, and in particular LSTMs, have shown success modeling the effect of truncated scales on the retained ones (i.e., model closure) with roots from the Mori-Zwanzig formalism. For example, Wang et al. \cite{wang2020recurrent} utilized a conditioned LSTM for the memory term in the GROM equations, representing the closure model, for parametric systems.} In addition to model imperfection, we also perturb the initial conditions to further mimic erroneous state estimates in practice. Meanwhile, we realize that, more often than not, sensor signals are noisy. So, we intentionally inject some noise to the synthesized observation data (using a twin-experiment approach). We test the performance of LSTM-N with various levels of measurement noises, initial field perturbations, and sensors signals sparsity.

\section{Wake-vortices and decay prediction system} \label{sec:wake}

Every aircraft generates a wake of turbulent air as it flies. This disturbance is caused by a pair of tornado-like counter-rotating vortices (called wake vortex) that trail from the tips of the wings \cite{spalart1998airplane}. Relatively turbulent weather conditions and rough terrain can help dissipate these vortices. A faster wake-decay was seen with increase in terrain roughness in \cite{tabib2016} for wind-farms, where it was observed that a secondary vortex (SV) gets established more rapidly around the periphery of primary wake-vortex (WV), and the subsequent interactions between SV and WV creates a higher turbulence state. The phenomena of WV rebound and generation of omega-shaped hair-pin vortices take place during this SV-WV interaction. This complex wake decay phenomena is also applicable for wake-vortex emanating from aircraft. Such facts have also been observed and exploited to artificially destroy wakes close to the ground using plates \cite{stephan2017numerical}. Therefore, understanding the complex dynamics of these wake vortices (WV) from its generation to decay is important in order to ensure flight safety, to increase airport capacity and to test new methods for destroying WVs and mitigating their effect.

\begin{figure}[ht]
\centering
\includegraphics[trim= 0 0 0 0, clip, width=0.95\textwidth]{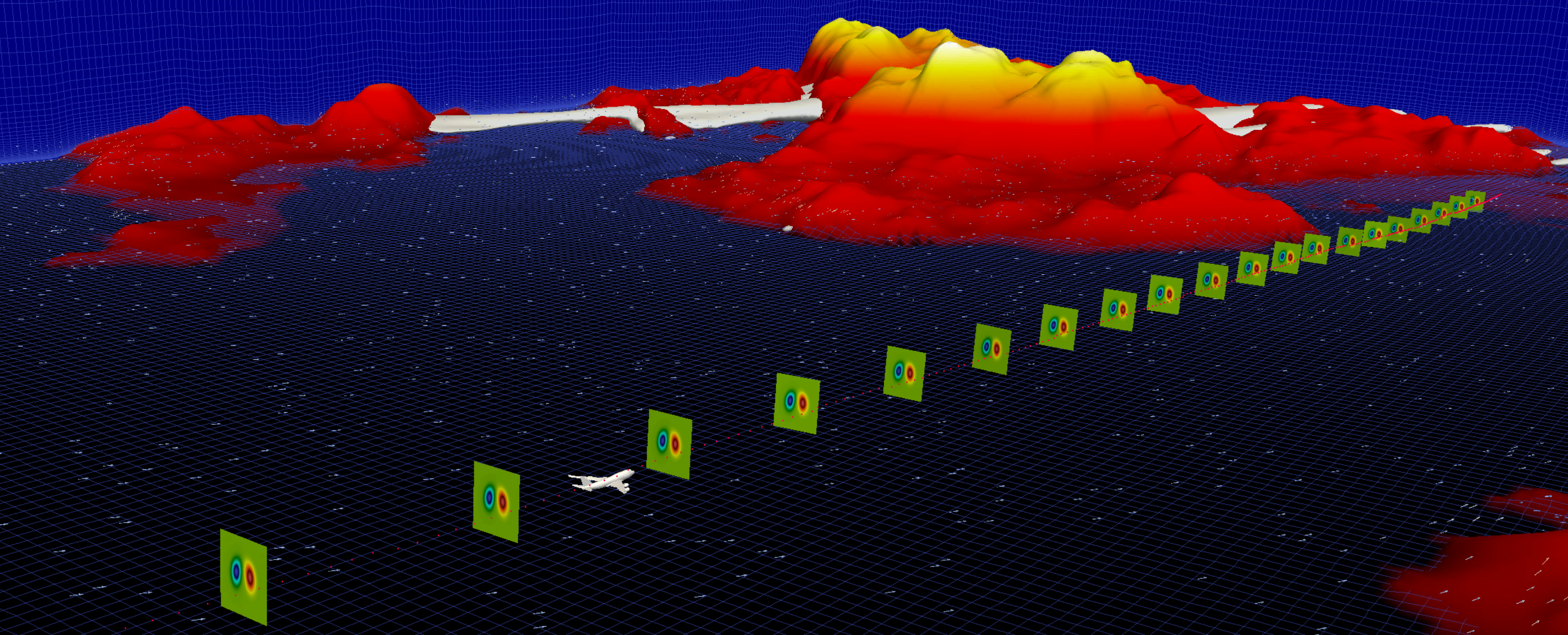}
\caption{Transported and diffused wakes on a set of 2D planes (a.k.a. gates) to make sure that the flight corridor is clear for the following aircraft.}
\label{fig:wake2}
\end{figure}

Air traffic control can potentially benefit from the emerging concept of a digital twin (DT), defined as the virtual replica of a physical system, where both of them are able to actively communicate with each other \cite{rasheed2020digital,ganguli2020digital,tao2018digital}. Given the WV and associated airport traffic case, a DT would receive streams of data, concerning operating conditions, airport traffic status, aircraft characteristics (e.g., weight, size) and flight mode (e.g., take-off or landing). Then, the DT should process these data and assess a bunch of possible scenarios corresponding to potential choices to provide an informed decision with regard to the separation distance and flight scheduling, for instance. Considering the WV problem, numerical modeling based on the Navier-Stokes equation, if accurate, can be a cost effective and easily employable pursuit for wake analysis. In Figure~\ref{fig:wake2}, an aircraft is admitted to land safely, based on the decay of wake-vortices from a leading aircraft. These wake-vortices are generated using the aircraft information and an analytical function on a set of two-dimensional (2D) planes (also called gates) perpendicular to the flight path \cite{fuchs2016wake}. Once the flight corridor is clear and free of any influence of the wake-vortices left behind by the leading aircraft, the following aircraft can land or take-off safely. 


Nonetheless, direct full order numerical simulations require large discretized systems for adequate approximation and are not practical for real-time wake prediction, which is an essential ingredient for feasible DT technologies. Therefore, reduced order modeling (ROM) rises as a natural choice for the successful implementation of DT applications. ROM represents a family of protocols that aim at emulating the relevant system's dynamics with minimal computational burden. Typical ROM approaches consist of two major steps; (1) tailor a low-order subspace, where the flow trajectory can be sufficiently approximated to live (see Section~\ref{sec:pod}), (2) build a surrogate model to cheaply propagate this trajectory in time (see Section~\ref{sec:gp}). Traditionally, building surrogate models to evolve on a reduced manifolds has relied on the projection of the full order model (FOM) operators onto a reduced subspace (e.g., using Galerkin-type techniques) to structure a reduced order model (ROM). Those FOM operators are usually the outcome of the numerical discretization of the well-established governing equation, derived from first principles and conservation laws. Such ROMs are attractive due to their reasonable interpretability and generalizability, as well as the existence of robust techniques for stability and uncertainty analysis. However, Galerkin ROM (GROM) can be expensive to solve for turbulent and advection-dominated flows. GROM also might suffer from inaccuracies and instabilities for long-time predictions. Meanwhile, in the digital twin context, the availability of rich stream of data and measurements opens new avenues for further ROM development. One way to utilize this abundance of data is the purely data-driven nonintrusive ROM (NIROM) approach. NIROMs have largely benefited from the widespread of open-source cutting edge and easy-to-use machine learning (ML) libraries, and cheap computational infrastructure to solely rely on data for building stable and accurate models, compared to their GROM counterparts \cite{kutz2017deep, brunton2019machine,brenner2019perspective, xie2019non, jian2019flowfield,pawar2019deep,san2019artificial,rahman2019nonintrusive,maulik2020time,renganathan2020machine,maulik2020probabilistic}. However, purely data-driven tools often lack human interpretability and generalizability, and sometimes become prohibitively ``data-hungry''.

Alternatively, hybrid approaches can be pursued, where data-driven tools only assist the physics-based models whenever data are available, rather than replacing them entirely. Data assimilation (DA) is a framework which can efficiently achieve this objective. DA generally refers to the discipline of intelligently fusing theory and observations to yield an optimal estimate of the system's evolution \cite{ghil1991data,kalnay2003atmospheric,lewis2006dynamic,lorenc1991meteorological,derber1989global}. Measurements are usually sparse (both in time and space) and noisy, while dynamical models are often imperfect due to the underlying assumptions and approximations introduced during either model derivation (e.g., neglecting source terms) or numerical solution of the resulting model (e.g., truncation error). DA algorithms have rich history in numerical weather predictions and are utilized on a daily basis to provide reliable forecasts. In this paper, we suppose that our dynamical model is the truncated GROM and we aim at utilizing live streams of measurements to correct the GROM trajectory. Specifically, we exploit the nudging method as our data assimilation framework, which works by relaxing the model state toward observations by adding correction (or nudging) terms, to mitigate the discrepancy between observations and model state \cite{di2020synchronization}. We employ LSTM mappings to account for this nudging term based on a combination between GROM predictions and available measurement data (see Section~\ref{sec:LSTM}). \textcolor{rev}{We highlight that a similar approach was proposed in \cite{pawar2020long} for the state correction of the Lorenz 96 system. However, in that study, no model order reduction was employed and the model was assumed to be perfect. Moreover, a static correction was adopted using distinct background and assimilated trajectories, where the background trajectory is not updated each assimilation window. In contrast, we follow a dynamic correction approach where the corrected states after each assimilation window act as the background initial condition for the following window. This overall methodology was briefly introduced in \cite{ahmed2020reduced} for the ROM of the 1D Burgers problem, and in the present study, we extend this approach to the 2D case of wake vortex transport problem and exploring higher sparsity ratios. We also aim at emphasizing and demonstrating the potential of the LSTM-N method for digital twin frameworks for real-time monitoring and control.}

\section{Methodology} \label{sec:rom}
In this section, we first give an overview of the full order model used to simulate the wake-vortex transport problem. Then, we present the reduced order formulations adopted in this study. In particular, we utilize proper orthogonal decomposition (POD) as a data-driven tool to extract the flow's coherent structures and build a reduced order subspace that best approximate the flow fields of interest. After that, we utilize a Galerkin approach to project the full order model operators onto that reduced space to build a \emph{structure-preserving}, \emph{physics-constrained} reduced order model.

\subsection{Vorticity Transport Equation} \label{sec:gov}
We consider the two-dimensional (2D) vorticity transport equation as our full order model (FOM) that resolves the wake-vortex transport, defined by the 2D Navier-Stokes equations in vorticity-streamfunction formulation as follows \cite{Gun89},
\begin{align} 
\dfrac{\partial \omega}{\partial t} + J(\omega,\psi) &= \dfrac{1}{\text{Re}} \nabla^2 \omega, \label{eq:NS}
\end{align}
where $\omega$ and $\psi$ denote the vorticity and streamfunction fields, respectively. $\text{Re}$ is the dimensionless Reynolds number, relating the inertial and viscous effects. Equation~\ref{eq:NS} is complemented by the kinematic relationship between the vorticity and streamfunction as below,
\begin{equation}
\nabla^2 \psi = -\omega. \label{eq:Poisson}
\end{equation}
Equation~\ref{eq:NS} and Equation~\ref{eq:Poisson} involve two operators, the Jacobian ($J(\cdot,\cdot)$) and the Laplacian ($\nabla^2 (\cdot)$) defined as
\begin{align}
    J(\omega,\psi) &= \dfrac{\partial \omega}{\partial x} \dfrac{\partial \psi}{\partial y} -  \dfrac{\partial \omega}{\partial y} \dfrac{\partial \psi}{\partial x}, \\
    \nabla^2 \omega &= \dfrac{\partial^2 \omega}{\partial x^2} + \dfrac{\partial^2 \omega}{\partial y^2}.
\end{align}

In order to mimic the wake-vortex problem, several models have been investigated \cite{gerz2002commercial,hallock2018review,ahmad2014review,holzapfel2003analysis}, including Gaussian vortex \cite{lugan2007simulation}, Rankine vortex \cite{rossow1977convective,aboelkassem2005viscous}, Lamb-Oseen vortex \cite{lamb1924hydrodynamics,holzapfel2000wake}, and Proctor vortex \cite{proctor1998nasa,proctor2000wake} among others. In the present study, we initialize the flow with a pair of counter-rotating Gaussian vortices with equal strengths centered at $(x_1,y_1)$ and $(x_2,y_2)$ as follows,
\begin{equation}
    \omega(x,y,0) =  \exp\left( -\rho \left[ (x-x_1)^2  + (y-y_1)^2 \right] \right) - \exp{\left( -\rho \left[ (x-x_2)^2 + (y-y_2)^2 \right] \right)}, \label{eq:init}
\end{equation}
where $\rho$ is an interacting parameter that controls the mutual interactions between the two vortical motions. \textcolor{rev}{We also consider periodic boundary conditions for the demonstration provided in the current study. }

\subsubsection{\textcolor{rev}{Numerical methods}}
\textcolor{rev}{For the spatial discretization of Eq.~\ref{eq:NS}, we use the standard second-order central finite difference scheme in linear term as follows,
\begin{equation}
    \nabla^2 \omega_{i,j} = \dfrac{\omega_{i+1,j} - 2\omega_{i,j} + \omega_{i-1,j} }{\Delta x^2} + \dfrac{\omega_{i,j+1} - 2\omega_{i,j} + \omega_{i,j-1} }{\Delta y^2}, \label{eq:FD1}
\end{equation}
where $\Delta x$ and $\Delta y$ are the mesh sizes in the $x$- and $y$-directions, respectively. For the nonlinear term, Arakawa\cite{arakawa1997computational} suggested that the conservation of energy, enstrophy, and skew-symmetry is sufficient to avoid computational instabilities stemming from nonlinear interactions. Therefore, we adopt the following second order Arakawa scheme for the Jacobian term,
\begin{equation}
    J(\omega_{i,j},\psi_{i,j}) = \dfrac{1}{3}(J_1 + J_2 + J_3),
\end{equation}
where the discrete Jacobians have the following forms,
\begin{align*}
    J_1 &= \dfrac{1}{4\Delta x \Delta y} \bigg[ (\omega_{i+1,j} - \omega_{i-1,j})(\psi_{i,j+1} - \psi_{i,j-1}) - (\omega_{i,j+1}-\omega_{i,j-1})(\psi_{i+1,j}-\psi_{i-1,j}) \bigg], \\
    J_2 &= \dfrac{1}{4\Delta x \Delta y} \bigg[ \omega_{i+1,j}(\psi_{i+1,j+1}-\psi_{i+1,j-1}) - \omega_{i-1,j}(\psi_{i-1,j+1}-\psi_{i-1,j-1}) \\
    &- \omega_{i,j+1}(\psi_{i+1,j+1}-\psi_{i-1,j+1}) + \omega_{i,j-1}(\psi_{i+1,j-1}-\psi_{i-1,j-1}) \bigg], \\
    J_3 &= \dfrac{1}{4\Delta x \Delta y} \bigg[ \omega_{i+1,j+1}(\psi_{i,j+1} - \psi_{i+1,j}) - \omega_{i-1,j-1}(\psi_{i-1,j}-\psi_{i,j-1}) \\
    &- \omega_{i-1,j+1}(\psi_{i,j+1}-\psi_{i-1,j}) + \omega_{i+1,j-1}(\psi_{i+1,j} - \psi_{i,j-1})  \bigg].
\end{align*}
}

\textcolor{rev}{Nonetheless, solving Eq.~\ref{eq:Poisson} is often the most computationally-demanding part for typical incompressible flow solvers. In our study, we make use of the periodicity of boundary conditions and implement a fast Poisson solver employing the fast Fourier transform (FFT). We first discretize Eq.~\ref{eq:Poisson} using the standard second-order central finite difference scheme (similar to Eq.~\ref{eq:FD1}) as follows,
\begin{equation}
   \dfrac{\psi_{i+1,j} - 2\psi_{i,j} + \psi_{i-1,j} }{\Delta x^2} + \dfrac{\psi_{i,j+1} - 2\psi_{i,j} + \psi_{i,j-1} }{\Delta y^2} =  -\omega_{i,j}, \label{eq:FD2}
\end{equation}
Then, we apply the forward FFT to find the Fourier coefficients $\hat{\omega}_{i,j}$ from the grid values of $\omega_{i,j}$. Thus, the Fourier coefficients $\hat{\psi}_{i,j}$ for the streamfunction can be evaluated as follows,
\begin{equation}
    \hat{\psi}_{i,j} = -\dfrac{\hat{\omega}_{i,j}}{c_1 \cos{(2\pi i/N_x)} + c_2 \cos{(2\pi j/N_y)} - c_3},
\end{equation}
where $c_1 = 2/\Delta x^2$, $c_2=2/\Delta y^2$, and $c_3 = c_1+c_2$. Finally, we apply an inverse FFT to find the grid values $\psi_{i,j}$ from their Fourier coefficients $\hat{\psi}_{i,j}$.}

\subsection{Proper Orthogonal Decomposition} \label{sec:pod}
The first step for building a projection-based reduced order model is to tailor a low-order subspace that is capable of capturing the essential features of the system of interest. In the fluid mechanics community, proper orthogonal decomposition (POD) is one of the most popular techniques in this regard \cite{taira2017modal,taira2020modal,rowley2017model}. Starting from a collection of system's realizations (snapshots), POD provides a systematic algorithm to construct a set of orthonormal basis functions (called POD modes) that best describes that collection of snapshot data (in the $\ell_2$ sense). More importantly, those bases are sorted based on their contributions to the system's total energy, making the modal selection a straightforward process. This is a significant advantage compared to other modal decomposition techniques like dynamic mode decomposition, where further sorting and selection criterion has to be carefully defined. Usually, the method of snapshots \cite{sirovich1987turbulence} is followed to perform POD efficiently and economically, especially for high dimensional systems. However, we demonstrate the singular value decomposition (SVD) based approach here for the sake of simplicity and brevity of presentation.

We collect $N$ system realizations, denoted as $\mathbf{\omega}_k = \{\omega(x_i,y_j, t_k)\}_{i=1,j=1,k=1}^{i=N_x,j=N_y,k=N}$, thus we build a snapshot matrix $\mathbf{A} \in \mathbb{R}^{M \times N}$ as $\mathbf{A} = [\mathbf{\Omega}_1, \mathbf{\Omega}_2, \dots, \mathbf{\Omega}_N]$, where $\mathbf{\Omega}_k \in \mathbb{R}^{M \times 1}$ is the $k^{th}$ snapshot reshaped into a column vector, $M$ is the number of spatial locations (i.e., $M=N_xN_y$), and $N$ is the number of snapshots. Then, a thin (reduced) SVD is performed on $\mathbf{A}$, 
\begin{equation} \label{eq:svd1}
    \mathbf{A} = \mathbf{U} \mathbf{\Sigma} \mathbf{V}^T,
\end{equation}
\textcolor{rev}{where $\mathbf{U} \in \mathbb{R}^{M \times N}$ is a matrix with orthonormal columns defining the left singular vectors of $\mathbf{A}$ while the columns of $\mathbf{V} \in \mathbb{R}^{N \times N}$ denote the right singular vectors of $\mathbf{A}$. We note that the columns of $\mathbf{U}$ represent the spatial basis, and the columns of $\mathbf{V}$ carry the respective temporal information.} The singular values of $\mathbf{A}$ are stored in descending order as the entries of the diagonal matrix $\mathbf{\Sigma} \in \mathbb{R}^{N \times N}$. For model order reduction purposes, only the first $R$ columns of $\mathbf{U}$, the first $R$ columns of $\mathbf{V}$, and the upper-left $R\times R$ sub-matrix of $\mathbf{\Sigma}$ are considered, corresponding to the largest $R$ singular values. Specifically, the first $R$ columns of $\mathbf{U}$ represent the most energetic $R$ POD modes, denoted as $\{\phi_k\}_{k=1}^{R}$ for now on.  

The vorticity field $\omega(x,y,t)$ is thus approximated as a linear superposition of the contributions of the first $R$ modes, which can be mathematically expressed using the Galerkin ansatz as
\begin{equation} \label{eq:uROM1}
    \omega(x,y,t) = \sum_{k=1}^{R} a_k(t) \phi_k(x,y),
\end{equation}
where $\phi_k(x,y)$ stand for the spatial modes, $a_k(t)$ designate the time-dependent modal coefficients/amplitudes (also known as generalized coordinates), and $R$ is the number of the retained modes in the ROM approximation (i.e., ROM dimension). We note that the POD basis functions $\phi$ are orthonormal by construction as follows,
\begin{equation}
\langle \phi_i ; \phi_j \rangle = 
     \begin{cases}
       1 &\quad\text{if } i = j \\
       0 &\quad\text{otherwise,}
     \end{cases}    
\end{equation}
where the angle parentheses $\langle \mathbf{\cdot} ; \mathbf{\cdot} \rangle$ stands for the standard inner product in Euclidean space (i.e., dot product). \textcolor{rev}{We highlight that the SVD-based computation of the POD basis imply the use of the Euclidean inner product. However, this might be problematic, especially when combined with non-uniform or unstructured grids. For such cases, the of other inner products (e.g., $L^2$ or $H^1$) would be recommended.}

Since the vorticity and streamfunction fields are related by Eq.~\ref{eq:Poisson}, they share the same modal amplitudes, $a_k(t)$. Moreover, the basis functions for the streamfunction (denoted as $\theta_k(x,y)$) are derived from those of the vorticity as follows,
\begin{equation}\label{eq:Poisson2}
\nabla^2 \theta_k = -\phi_k, \quad \quad k=1,2, \dots, R,
\end{equation}
and the ROM approximation of the streamfunction field can be written as
\begin{equation} \label{eq:uROM2}
    \psi(x,y,t) = \sum_{k=1}^{R} a_k(t) \theta_k(x,y),
\end{equation}

\subsection{Galerkin Projection} \label{sec:gp}

After constructing a set of POD basis functions, an orthogonal Galerkin projection can be performed to obtain the Galerkin ROM (GROM). To do so, the ROM approximation (Eq.~\ref{eq:uROM1}-\ref{eq:uROM2}) is substituted into the governing equation (Eq.~\ref{eq:NS}). Noting that the POD bases are only spatial functions (i.e., independent of time) and the modal amplitudes are independent of space, we get the the following set of ordinary differential equations (ODEs) representing the tensorial GROM
\begin{align}
    \dfrac{\text{d}a_k}{\text{d}t} &= \sum_{i=1}^{R} \mathfrak{L}_{i,k} a_i + \sum_{i=1}^{R} \sum_{j=1}^{R} \mathfrak{N}_{i,j,k} a_i a_j, \label{eq:rom}
\end{align}
where $\mathfrak{L}$ and $\mathfrak{N}$ are the matrix and tensor of predetermined model coefficients corresponding to linear and nonlinear terms, respectively. Those can be precomputed once during an offline stage as
\begin{align*}
    \mathfrak{L}_{i,k} = \big\langle  \dfrac{1}{\text{Re}} \nabla^2 \phi_i ; \phi_k \big\rangle, \qquad 
    \mathfrak{N}_{i,j,k} = \big\langle -J(\phi_i,\theta_j) ; \phi_k \big\rangle.
\end{align*}
Equation~\ref{eq:rom} can be rewritten in compact form as
\begin{equation} \label{eq:romcon}
    \dot{\mathbf{a}} = \mathbf{f}(\mathbf{a}),
\end{equation}
where the (continuous-time) model map $\mathbf{f}$ is defined as follows,
\begin{equation*}
\mathbf{f} = 
\begin{bmatrix} 
    \sum_{i=1}^{R} \mathfrak{L}_{i,1} a_i + \sum_{i=1}^{R} \sum_{j=1}^{R} \mathfrak{N}_{i,j,1} a_i a_j \\
    \sum_{i=1}^{R} \mathfrak{L}_{i,2} a_i + \sum_{i=1}^{R} \sum_{j=1}^{R} \mathfrak{N}_{i,j,2} a_i a_j \\
    \vdots \\
    \sum_{i=1}^{R} \mathfrak{L}_{i,R} a_i + \sum_{i=1}^{R} \sum_{j=1}^{R} \mathfrak{N}_{i,j,R} a_i a_j
\end{bmatrix}.
\end{equation*}
Alternatively, Eq.~\ref{eq:romcon} can be given in discrete-time form as 
\begin{equation} \label{eq:romdis}
    \mathbf{a}^{n+1} = \mathbf{M}(\mathbf{a}^n),
\end{equation}
where $\mathbf{M}(\cdot)$ is \textcolor{rev}{a one time-step forward mapping} obtained by any suitable temporal integration technique. Here, we use the fourth-order Runge-Kutta (RK4) method as follows,
\begin{align}
    \mathbf{a}^{n+1} &= \mathbf{a}^n + \dfrac{\Delta t}{6} (\mathbf{g}_1 + 2\mathbf{g}_2 + 2\mathbf{g}_3 + \mathbf{g}_4),
\end{align}
where 
\begin{align*}
    \mathbf{g}_1 = \mathbf{f}(\mathbf{a}^n), \quad  \mathbf{g}_2 = \mathbf{f}(\mathbf{a}^n + \dfrac{\Delta t}{2} \mathbf{g}_1), \quad    \mathbf{g}_3 = \mathbf{f}(\mathbf{a}^n + \dfrac{\Delta t}{2}  \mathbf{g}_2), \quad   \mathbf{g}_4 = \mathbf{f}(\mathbf{a}^n + \Delta t \mathbf{g}_3).
\end{align*}
Thus the discrete-time map defining the transition from time $t_n$ to time $t_{n+1}$ is written as
\begin{align}
\mathbf{M}(\mathbf{a}^n) = \mathbf{a}^n + \dfrac{\Delta t}{6} (\mathbf{g}_1 + 2\mathbf{g}_2 + 2\mathbf{g}_3 + \mathbf{g}_4). 
\end{align}

\subsection{Long Short-Term Memory Nudging} \label{sec:LSTM}
Due to the quadratic nonlinearity in the governing equation (and consequently the GROM), the \emph{online} computational cost of solving Eq.~\ref{eq:rom} is $O(R^3)$ (i.e., it scales cubically with the number of retained modes). Therefore, this has to be kept as low as possible for feasible implementation of ROM in digital twin applications that require near real-time responses. However, this is often not an easy task for systems with slow decay of the Kolmogorov $n$-width. Examples include advection-dominated flows with strong nonlinear interactions among a wide span of modes. As a result, the resulting GROM is intrinsically imperfect model. \textcolor{rev}{This imperfection implies that the GROM might give inaccurate or false predictions even when fed with the true initial conditions and in the absence of numerical discretization errors.}

Moreover, in most realistic cases, proper specification of the initial state, boundary conditions, and/or model parameters is rarely attainable. This uncertainty in problem definition, in conjunction with model imperfection, poses challenges for accurate predictions. In this study, we put forth a nudging-based methodology that fuses prior model forecast (using imperfect initial condition specification and imperfect model) with the available Eulerian sensor measurements to provide more accurate posterior prediction. Relating our setting to realistic applications, we build our framework on the assumption that measurements are noisy and sparse both in space and time. Nudging has a prestigious history in data assimilation, being a simple and unbiased approach. The idea behind nudging is to penalize the dynamical model evolution with the discrepancy between model's predictions and observations \cite{anthes1974data,lei2015nudging,stauffer1993optimal}. In other words, the forward model given in Eq.~\ref{eq:romdis} is supplied with a nudging (or correction) term rewritten in the following form, 
\begin{equation}\label{eq:nudge1}
    \mathbf{a}^{n+1} = \mathbf{M}(\mathbf{a}^n) + \mathbf{G}(\mathbf{z}^{n+1}-h(\mathbf{a}^{n+1})),
\end{equation}
where $\mathbf{G}$ is called the nudging (gain) matrix and $\mathbf{z}$ is the set of measurements (observations), while $h(\cdot)$ is a mapping from model space to observation space. For example, $h(\cdot)$ can be a reconstruction map, from ROM space to FOM space. In other words, $h(\mathbf{a})$ represents the ``model forecast for the measured quantity'', while $\mathbf{z}$ is the ``actual'' observations. Despite the simplicity of Eq.~\ref{eq:nudge1}, the specification/definition of the gain matrix $\mathbf{G}$ is not as simple \cite{zou1992optimal,vidard2003determination,auroux2005back,lakshmivarahan2013nudging}.

In the proposed framework, we utilize a recurrent neural network, namely the long short-term memory (LSTM) variant, to define the nudging map. We denote our approach as LSTM-Nudging (LSTM-N). In particular, Eq.~\ref{eq:nudge1} implies that each component of $\mathbf{a}^{n+1}$ (i.e., $a_1,a_2\dots,a_R$) is corrected using a \emph{linear} superposition of the components of $\mathbf{z}^{n+1}-h(\mathbf{a}^{n+1})$, weighted by the gain matrix. Here, we relax this linearity assumption and generalize it to a possibly nonlinear mapping $\mathbf{C}(\mathbf{a}, \mathbf{z})$ as,
\begin{equation}\label{eq:nudge2}
    \mathbf{a}^{n+1} = \mathbf{M}(\mathbf{a}^n) + \mathbf{C}(\mathbf{a}_b^{n+1}, \mathbf{z}^{n+1}),
\end{equation}
where the map $\mathbf{C}(\mathbf{a}, \mathbf{z})$ is learnt (or inferred) using an LSTM neural network, and $\mathbf{a}_b^{n+1}$ is the prior model prediction computed using imperfect model and/or imperfect initial conditions (called background in data assimilation terminology), defined as $\mathbf{a}_b^{n+1} = \mathbf{M}(\mathbf{a}^n)$. Thus, Eq.~\ref{eq:nudge2} can be rewritten as follows,
\begin{equation}\label{eq:nudge3}
    \mathbf{a}^{n+1} = \mathbf{a}_b^{n+1} + \mathbf{C}(\mathbf{a}_b^{n+1}, \mathbf{z}^{n+1}).
\end{equation}

In order to learn the map $\mathbf{C}(\mathbf{a}_b,\mathbf{z})$, we consider the case with imperfect model, defective initial conditions, and noisy observations. Moreover, we suppose sensors are sparse in space and measurement signals are sparse in time, too. Specifically, we use sensors located at a few equally-spaced grid points, but a generalization to off-grid or adaptive sensor placement is possible. Also, we assume sensors send measurement signals every $\tau$ time units. In order to mimic sensor measurements and noisy initial conditions, we run a twin experiment as follows,
\begin{enumerate}
    \item Solve the FOM equation (i.e., Eq.~\ref{eq:NS}) and sample \emph{true} field data ($\omega_{true}(x,y,t_n)$) each $\tau$ time units. In other words, store $\omega_{true}(x,y,t_n)$ at $t_n\in \{0,\tau,2\tau,\dots T\}$ where $T$ is the total (maximum) time and $\tau$ is the time window over which measurements are collected.
    \item Define erroneous initial field estimate as $\omega_{err} (x,y,t_n) = \omega_{true}(x,y,t_n) + \epsilon_b$, where $t_n\in \{0,\tau,2\tau,\dots T-\tau\}$. Here, $\epsilon_b$ stands for noise in initial state estimate, assumed as white Gaussian noise with zero mean and covariance matrix $B$ (i.e., $\epsilon_b \sim {\cal{N}}(0,B)$).
    \item Define sparse and noisy measurements as $\mathbf{z} = \omega_{true}(x_{Obs},y_{Obs},t_n) + \epsilon_m $, for $t_n\in \{\tau,2\tau,\dots T\}$. Similarly, $\epsilon_m$ stands for the measurements noise, assumed to be white Gaussian noise with zero mean and covariance matrix $Q$ (i.e., $\epsilon_m \sim {\cal{N}}(0,Q)$).
\end{enumerate}
For LSTM training data, we project the erroneous field estimates (from Step 2) onto the POD basis functions to get the erroneous POD modal coefficients (i.e., $\mathbf{a}_{err}(t_n)$, for $t_n\in \{0,\tau,2\tau,\dots T-\tau\}$. Then, we integrate those erroneous coefficients for $\tau$ time units to get the background prediction $\mathbf{a}_b(t_n)$, for $t_n\in \{\tau,2\tau,\dots T\}$. \textcolor{rev}{We note that in the actual deployment, the ROM solver is \emph{re-initialized} with the nudged states each $\tau$ time units when measurements become available. Therefore, we perform our training on a bunch of samples initiated at time $t_n$ (with the Gaussian noise representing the uncertainty) and integrated up to $t_n+\tau$ to mimic the re-initialization each $\tau$ time units. That said, in Step 2, we use the same level of uncertainty at $t=0$ (defined by the background covariance matrix $B$) as an upper limit for the uncertainty at the beginning of the following periods (i.e., $t=\tau, 2\tau, \dots$) since the nudging algorithm does not provide an update for the background covariance matrix (unlike Kalman filtering approaches which evolve this information along with the state).}

We train the LSTM using $\mathbf{a}_b(t_n)$ and $\mathbf{z}(t_n)$ as inputs, and set the target as the correction $(\mathbf{a}_{true}(t_n) - \mathbf{a}_b(t_n))$, for $t_n\in \{\tau,2\tau,\dots T\}$. The true modal coefficients ($\mathbf{a}_{true}$) are obtained by projecting the \emph{true} field data (from Step 1) onto the POD bases, where the projection is defined via the inner product as $a_k(t) = \langle \omega(x,y,t) ; \phi_k(x,y) \rangle$. \textcolor{rev}{We remark that the LSTM-N framework is fed with the whole $\mathbf{a}_b(t_n)$ vector (not individual components) after propagating the GROM in time. Therefore, it is independent of the numerical method of time integration and it is generalizable to either explicit or implicit schemes.} In order to enrich the training data set, Step 2 and Step 3 are repeated several times giving an ensemble of erroneous state estimates and noisy measurements at every time instant of interest. Each member of those ensembles represents one training sample. This also assists the LSTM network to handle wider range of noise.

\textcolor{rev}{We should highlight here that extra care should be taken when considering data-driven correction approaches as data-driven closure might yield non-physical results, and putting some extra constraints on the characteristics of the predicted closure is essential in several cases. Wu et al. \cite{wu2019reynolds} demonstrated that the Reynolds-averaged Navier–Stokes (RANS) with explicit treatment of data-driven closure can be ill-conditioned and yield large errors. In another study, Wu et al. \cite{wu2018physics} also proposed a physics-informed machine learning approach for improved data-driven RANS closure by training the ML models for the linear and nonlinear parts of the Reynolds stress separately.} Nonetheless, we emphasize that the proposed LSTM-N approach not only cures model imperfection (i.e., provides model closure as well as accounts for any missing physical processes), but also treats uncertainties in initial state estimates. This might be caused by the selection of inaccurate wake vortex model, or the idealizations embedded in this model compared to reality. Moreover, the field measurements (i.e., the nudging data) are assumed to be sparse and noisy to mimic real-life situations. 

\section{Results} \label{sec:res}
\begin{figure}[ht]
\centering
\includegraphics[trim= 0 0 0 0, clip, width=0.95\textwidth]{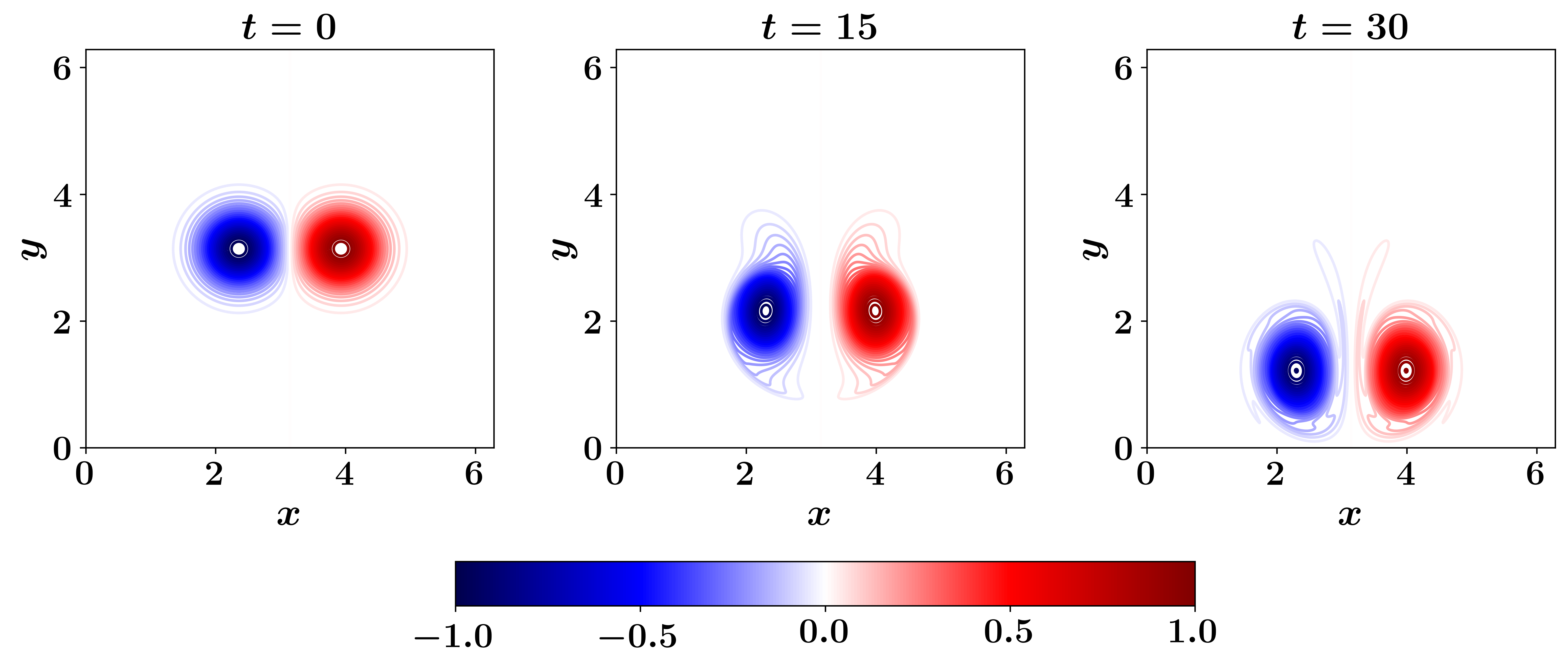}
\caption{Evolution of the FOM vorticity field for the wake vortex transport problem with a Reynolds number of $10^4$. Flow is initiated at time $t=0$ with a pair of Gaussian distributed vortices.}
\label{fig:FOM}
\end{figure}

In order to test and verify the proposed ideas, we consider a square 2D domain with a side length of $2\pi$. Flow is initiated using a pair of Gaussian vortices as given in Eq.~\ref{eq:init} centered at  $(x_1,y_1) = \left( \dfrac{5\pi}{4},\pi \right)$ and $(x_2,y_2) = \left( \dfrac{3\pi}{4},\pi \right)$ with an interaction parameter of $\rho = \pi$. Results in this section are shown at $Re=10^4$. For FOM simulations, a regular Cartesian grid resolution of $512\times512$ is considered (i.e., $\Delta x = \Delta y = 2\pi/512$), with a time-step of $0.001$. Snapshots of vorticity fields are collected every 100 time-steps for $t\in [0,30]$, totalling 300 snapshots. The evolution of the wake vortex problem is depicted in Figure~\ref{fig:FOM}, demonstrating the convective and interactive mechanisms affecting the transport and development of the two vortices.

For ROM computations, 6 modes are retained in the reduced order approximation (i.e.,  $R=6$), \textcolor{rev}{capturing more than $99\%$ of the snapshot data variance.} A time step of $0.1$ is adopted for the temporal integration of GROM equations. In order to implement the LSTM-N approach, we begin at erroneous initial condition defined as $\omega_{err}(x,y,0) = \omega_{true}(x,y,0) + \epsilon_b$, where $\omega_{true}(x,y,0)$ is defined with Eq.~\ref{eq:init}, and $\epsilon_b$ is a white Gaussian noise with zero mean and covariance matrix $B$. For simplicity, we assume $B=\sigma_b^2 \mathbf{I}$, where $\sigma_b$ is the standard deviation in the ``background'' estimate of the initial condition and $\mathbf{I}$ is the identity matrix. We note that this formulation implies that the errors in our estimates of the initial vorticity field at different spatial locations are uncorrelated. As nudging field data, we locate sensors to measure the vorticity field $\omega(x,y,t)$ every 64 grid points (i.e., 9 sensors in each direction, with $s_{freq} = 64$, where $s_{freq}$ is the number of spatial steps between sensors locations), and collect measurements every 10 time steps (i.e., each $1$ time unit with $t_{freq}=10$, where $t_{freq}$ is the number of time steps between measurement signals). To account for noisy observations, a white Gaussian noise of zero mean and covariance matrix of $Q$ is added to the true vorticity field obtained from the FOM simulation at sensors locations. Similar to $B$, we set $Q=\sigma_m^2 \mathbf{I}$, where $\sigma_m$ is the standard deviation of measurement noise. This assumes that the noise in sensors measurements are not correlated to each other, and all sensors have similar quality (i.e., add similar amounts of noise to the measurements). As a base case, we set $\sigma_b=1.0$, and $\sigma_m=1.0$. \textcolor{rev}{These levels of initial condition perturbation and measurement noise are guided by the values vorticity fields varying between $-1.0$ and $1.0$ as indicated by the colorbar in Figure~\ref{fig:FOM}. Thus, setting $\sigma_b=1.0$, and $\sigma_m=1.0$ guarantees extensive levels of uncertainty in the background information and collected observations.}

\begin{figure}[ht]
\centering
\includegraphics[width=0.95\textwidth]{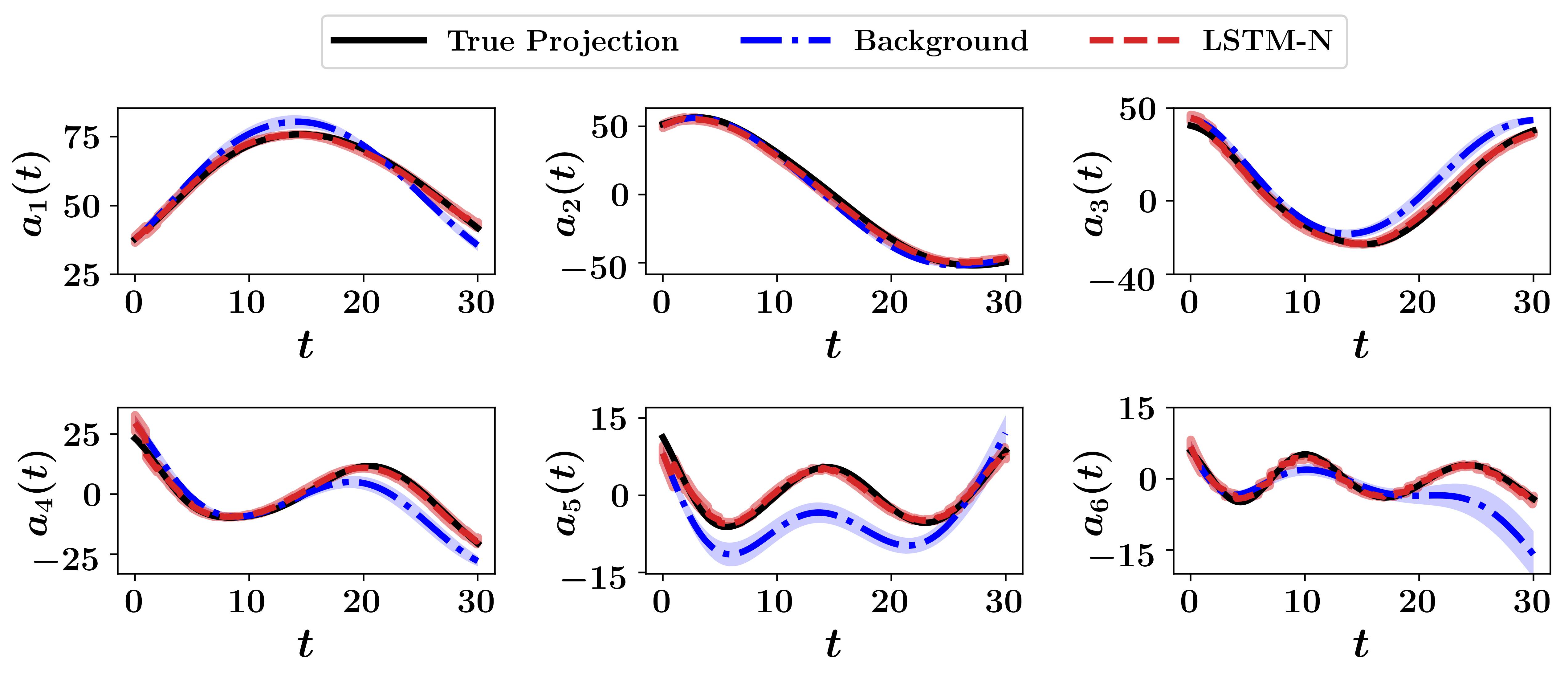}
\caption{Temporal evolution of the POD modal coefficients for the 2D wake vortex transport problem. \textcolor{rev}{The dark lines denote the mean value from a sample of 30 realizations (using different seeds for the random number generator to simulate initial condition perturbation and measurement noise). The shaded area defines the mean values +/- the standard deviation of the sample.} [Base case with $\sigma_b=1.0$ and $\sigma_m=1.0$]}
\label{fig:a}
\end{figure}

The procedure presented in Sec.~\ref{sec:LSTM} is applied using the numerical setup described above, and compared against the reference case of GROM with the erroneous initial condition and inherent model imperfections due to modal truncation (denoted as background forecast). \textcolor{rev}{Since we synthesize the initial condition perturbation and measurement noise using a pseudo-random number generator, we utilize an ensemble of 30 realizations with different seed numbers and the sample mean is computed from these 30 experiments. Whenever applicable, we also sketch uncertainty regions bounded by the sample mean +/- the standard deviation.} In Figure~\ref{fig:a}, the temporal evolution of the POD modal coefficients is shown for the true projection, background, and LSTM-N results. The true projection results are obtained by the projection of the true FOM field at different time instants onto the corresponding basis functions (i.e., via inner product, $a_{k,true}(t) = \langle \omega(x,y,t) ; \phi_k(x,y) \rangle$). The background trajectory is the reference solution obtained by standard GROM using the erroneous initial condition, without any closure or corrections. It can be seen that the background trajectory gets off the true trajectory by time as a manifestation of model. Also, note that the background solution does not begin from the same point as true projection due to the noise in the initial condition. On the other hand, the LSTM-N yields very good predictions, comparable to the true projection solution, implying that the approach is capable of blending noisy observations with a prior estimate to gain more accurate predictions.

\begin{figure}[ht]
\centering
\includegraphics[width=0.95\textwidth]{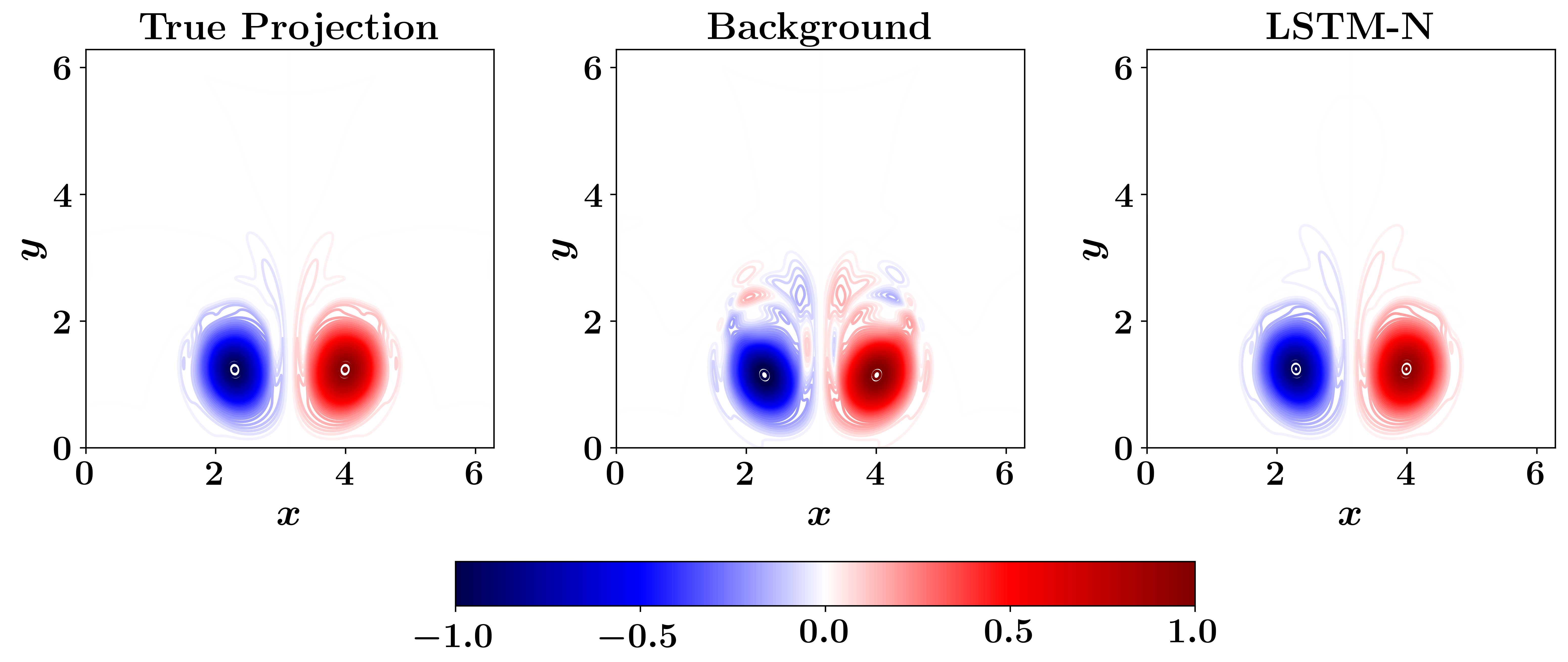}
\caption{Final vorticity field (at $t=30$) for the wake-vortex transport problem, with $\sigma_b=1.0$, and $\sigma_m=1.0$. \textcolor{rev}{Results show the mean of an ensemble of 30 realizations with different seeds for the random number generator.}}
\label{fig:u}
\end{figure}

In order to better visualize the predictive capabilities of the LSTM-N methodology, we compute the reconstructed vorticity field using Eq.~\ref{eq:uROM1}. \textcolor{rev}{The ensemble average of final field reconstruction (at $t=30$) is shown in Figure~\ref{fig:u}, comparing the true projection, background, LSTM-N results.} Note that the field obtained from true projection at any time instant can be computed as $\omega_{true}(x,y,t) = \sum_{k=1}^{R} a_{k,true}(t) \phi_k(x,y)$, and represents the optimal reduced-rank approximation that can be obtained using a linear subspace spanned by $R$ bases. Comparing true projection results from Figure~\ref{fig:u} against FOM at final time from Figure~\ref{fig:FOM} reveals that, for this particular case, 6 modes are qualitatively capable to capture most of the relevant features of the flow field. The LSTM-N outputs significantly match the projection of the FOM field, while the background forecasts show some visible deviations.

\textcolor{rev}{We also compare the LSTM-N results against a simple forward nudging implementation (denoted as linear nudging) \cite{asch2016data}, where Eq.~\ref{eq:nudge1} is rewritten as follows,
\begin{equation}
    \mathbf{a}^{n+1} = \mathbf{M}(\mathbf{a}^n) + \zeta \mathbf{D}_h^T (\mathbf{z}^{n+1}-h(\mathbf{a}^{n+1})),\label{eq:linNudge}
\end{equation}
where $\zeta$ is a constant and $\mathbf{D}_h$ is the Jacobian matrix of the observation operator $h(\cdot)$. In the present study, we find that a value of $\zeta=\Delta t$ provides acceptable results. We also highlight that different data assimilation techniques might be adopted. For example, in the three dimensional variational (3DVAR) framework \cite{lewis2006dynamic}, solving Eq.~\ref{eq:nudge1} and defining the gain matrix are reformulated as an optimization problem. In particular, the output of the data assimilation (called analysis or analyzed state) is defined as the minimizer of the following cost functional, 
\begin{equation}
    J(\mathbf{a}^{n+1}) = \|\mathbf{a}^{n+1} - \mathbf{M}(\mathbf{a}^n) \|_{W_1}^2 + \| \mathbf{z}^{n+1} - h(\mathbf{a}^{n+1})  \|_{W_2}^2,
\end{equation}
where ${W_1}$ and ${W_2}$ are some suitable symmetric positive definite (SPD) matrix. Most often, ${W_1}$ is set as the background covariance matrix, while $W_2$ is defined using the measurement noise covariance matrix. The 3DVAR, as a variational approach, addresses the inference problem by formulating an optimization problem, taking into account measurement noise and background perturbation. For linear observation operator $h(\cdot)$, this can be directly solved with proper definitions of the weighting matrices. However, for nonlinear operators, some approximations and/or linearization become necessary. Indeed, the LSTM-N can be also thought of as minimizing that cost functional iteratively using the training samples and the back-propagation algorithm.}

\begin{figure}[ht]
\centering
\includegraphics[width=0.95\textwidth]{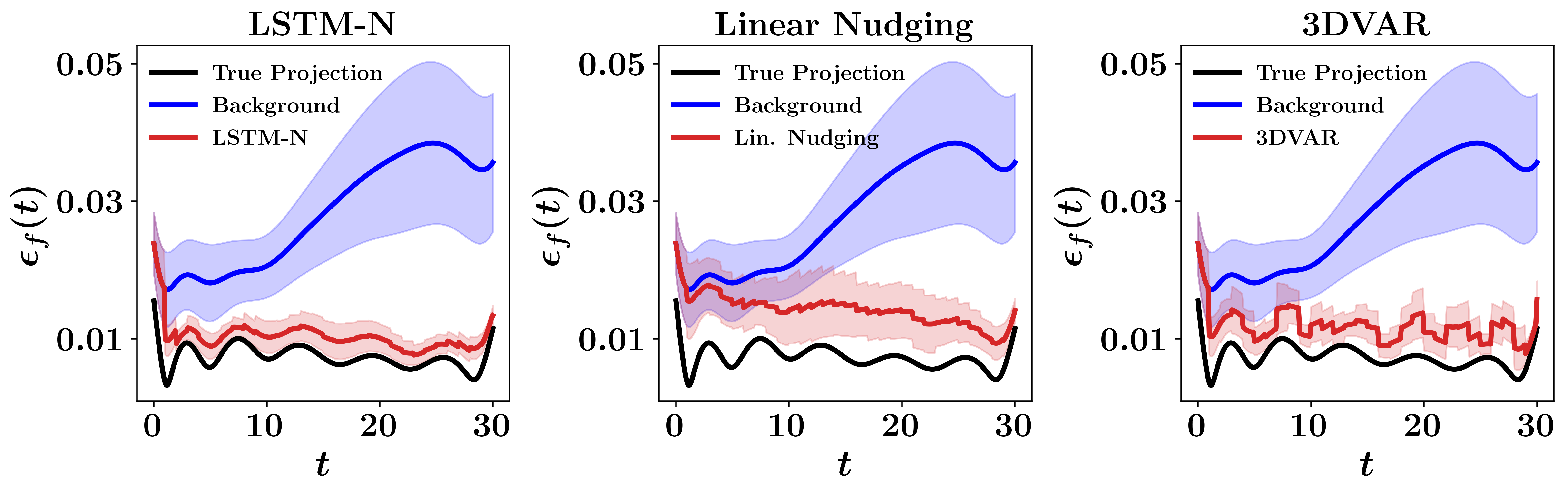}
\caption{\textcolor{rev}{Root mean-squares error for the wake-vortex transport problem, with $\sigma_b=1.0$, and $\sigma_m=1.0$, using the LSTM-N, Linear Nudging, and 3DVAR approaches.}}
\label{fig:rmse}
\end{figure}

For quantitative assessment, the root mean-squares error (RMSE) of the \emph{whole} reconstructed field with respect to the FOM solution is calculated as a function of time as follows,
\begin{equation} \label{eq:RMSE}
    \epsilon_f(t) = \sqrt{\dfrac{1}{N_x N_y}\sum_{i=1}^{N_x} \sum_{j=1}^{N_y}{\bigg(\omega_{FOM}(x_i,y_j, t) - \omega_{ROM}(x_i, y_j,t) \bigg)^2} },
\end{equation}
where $\omega_{FOM}$ is the true vorticity field obtained from solving the FOM equation, while $\omega_{ROM}$ is the reduced order approximation computed through true projection, background (reference) solution, or LSTM-N method. \textcolor{rev}{The RMSE at different times is plotted in Figure~\ref{fig:rmse}, comparing the LSTM-N, Linear Nudging, and 3DVAR results.  demonstrating the capability of LSTM-N framework to efficiently recover the optimal reconstruction given a few sparse measurements. We also remark that we find that the performance of the Linear Nudging and the 3DVAR approaches is highly dependent on the level of measurement noise. Indeed, for larger values of $\sigma_m$, the solution does not converge. This behavior might be attributed to the definition of the observation map using the ROM reconstruction process. In particular, we find that the basis functions, when sampled at sparse locations, yield ill-conditioned matrices, and the resulting mapping becomes very sensitive to the measurement noise. One approach to mitigate this issue is to exploit compressed sensing and adaptive sampling techniques to improve this reconstruction mapping and minimize the effect of noise.}

\begin{figure}[ht]
\centering
\includegraphics[width=0.95\textwidth]{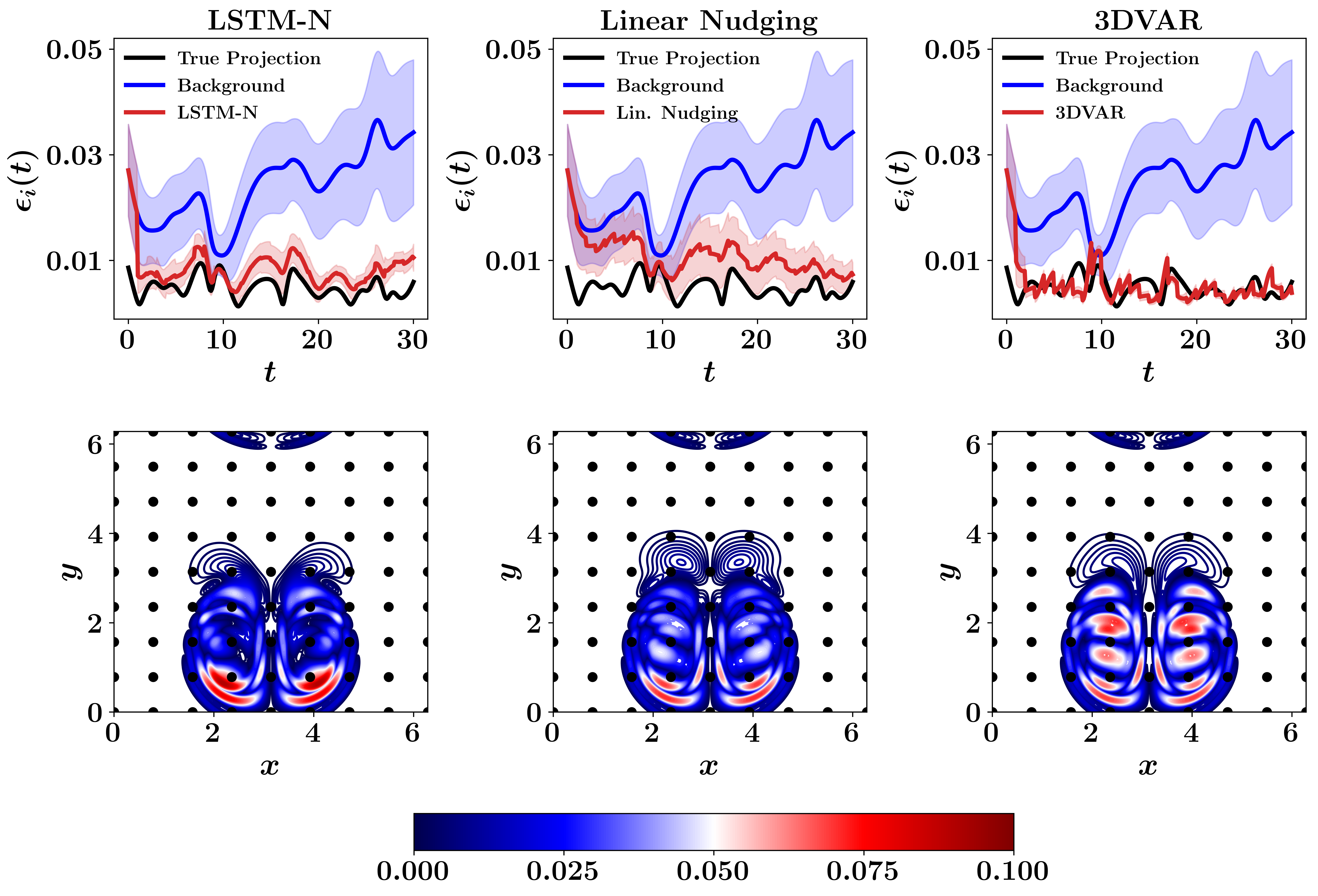}
\caption{\textcolor{rev}{Comparison of LSTM-N, Linear Nudging, and 3DVAR predictions, with $\sigma_b=1.0$, and $\sigma_m=1.0$. Top rows depicts the inference quality metric define by Eq.~\ref{eq:inf} and bottom row shows the absolute error between FOM solution and the ensemble mean for the respective approaches. Observation locations are denoted using black circles.}}
\label{fig:comp}
\end{figure}

\textcolor{rev}{Another metric to reveal the inference quality of the framework is defined using the error between the true measurements and the predicted solution at the sensors locations only as follows,
\begin{equation} \label{eq:inf}
    \epsilon_i(t) = \sqrt{\dfrac{1}{N_x^{meas} N_y^{meas}}\sum_{i=1}^{N_x^{meas}} \sum_{j=1}^{N_y^{meas}}{\bigg(\omega_{FOM}(x_i,y_j, t) - \omega_{ROM}(x_i, y_j,t) \bigg)^2} },
\end{equation}
where $N_x^{meas}$ and $N_y^{meas}$ represent the number of sensors in the $x$- and $y$-directions, respectively. Results for LSTM-N, Linear Nudging, and 3DVAR are presented in Figure~\ref{fig:comp} (top row). We observe that the 3DVAR gives better accuracy at observation locations than the nudging frameworks (LSTM-N and Linear Nudging). In order to understand this observation, we plot the absolute error between FOM solution and the ensemble average at final time from LSTM-N, Linear Nudging, and 3DVAR (i.e., $|\omega_{FOM}(x_i,y_j, t) - \omega_{ROM}(x_i, y_j,t)|$) in Figure~\ref{fig:comp} (bottom row). We can notice that the 3DVAR error at the observation locations (denoted as black circles) is small (note the bluish color). However, away from these measurement points, the error becomes larger (as denoted by the reddish color). On the other hand, for the LSTM-N, the error is small almost everywhere except for a small portion near the front of the vortices.
}

\subsection{Effect of Noise} \label{sec:noise}
\begin{figure}[ht]	
	\centering
	\includegraphics[width=1\linewidth]{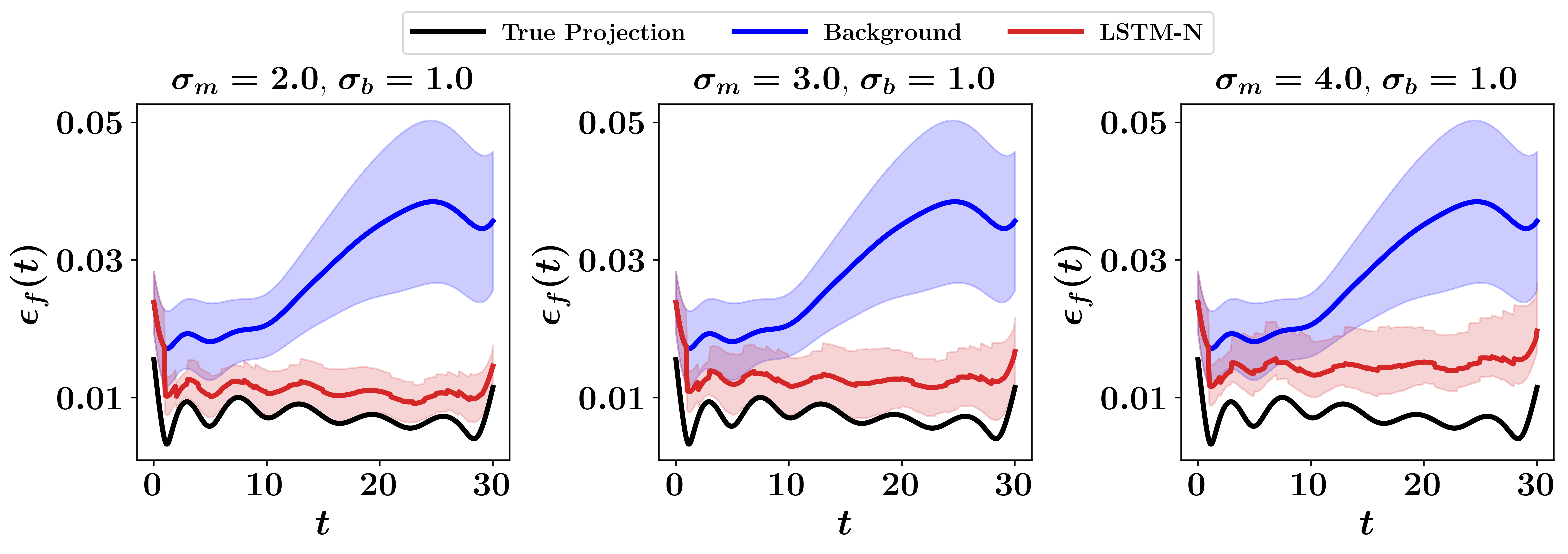}
	\caption{Root mean-squares error $\epsilon_f$ of LSTM-N predictions for different levels of measurement noise.} \label{fig:mnoise}
\end{figure}

Next, We explore the effect of noise on the LSTM-N results. In other words, we investigate how much noise the framework can tolerate. We note that we keep the same LSTM, trained with the base level of noise (i.e., $\sigma_b=1.0$ and $\sigma_m=1.0$) while we test it using different levels of noise. First, we gradually increase the standard deviation of measurement noise from $1.0$ to \textcolor{rev}{$2.0$ (2 times larger), $3.0$ (3 times larger), and $4.0$ (4 times larger)}. \textcolor{rev}{In Figure~\ref{fig:mnoise}, we plot the temporal evolution of $\epsilon_f$ metrics for the explored levels of measurement noise. We find that performance deteriorates a bit with an increase in measurement noise, especially in terms of uncertainty levels. Nonetheless, the predicted results are still significantly better than the background forecast (starting from the same initial conditions).}

\begin{figure}[ht]	
	\centering
	\includegraphics[width=1\linewidth]{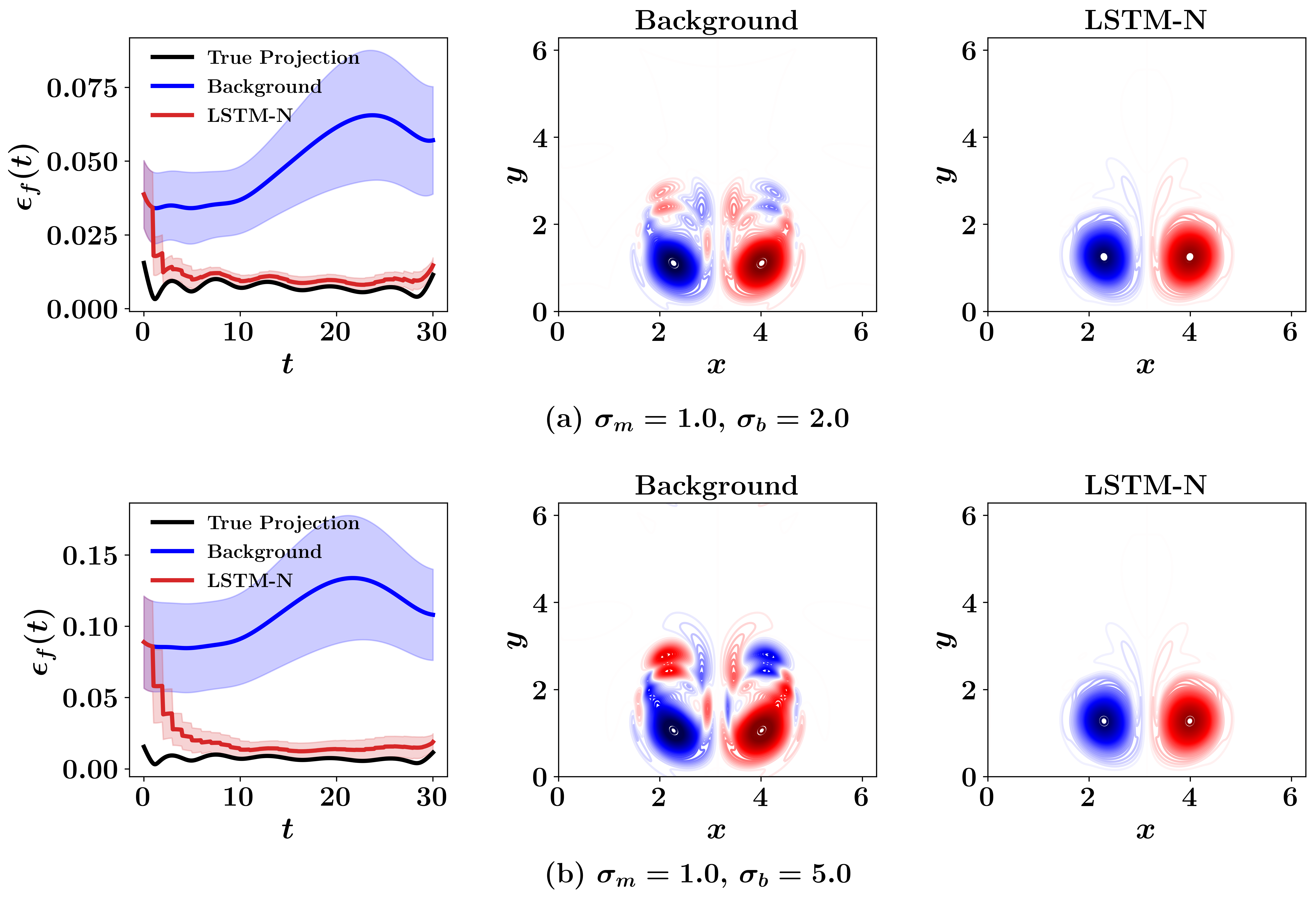}
	\caption{Reconstructed vorticity fields at final time, along with $RMSE$ for different levels of background noise.} \label{fig:bnoise}
\end{figure}

For testing the effect of initial state perturbation, we increase $\sigma_b$ from $1$ to $2$, and $5$. Figure~\ref{fig:bnoise} displays the effect of those levels of initial field perturbations on background forecasts. Despite that, LSTM-N is performing very well even at those high levels of initial perturbations. This is even clearer from the $RMSE$ plots, beginning from relatively large values and quickly decaying to the level of true projection once measurements are available. From Figure~\ref{fig:mnoise} and Figure~\ref{fig:bnoise}, we can deduce that the influence of the level of measurement noise on LSTM-N performance is more prominent that of the initial field perturbation. We reiterate that in both cases, the LSTM is trained with $\sigma_b=1.0$ and $\sigma_m=1.0$ and tested for different noise and perturbation levels.

\subsection{Effect of Measurements Sparsity} \label{sec:sparse}
\begin{figure}[ht]	
	\centering
	\includegraphics[trim= 0 0 0 0, clip, width=0.95\linewidth]{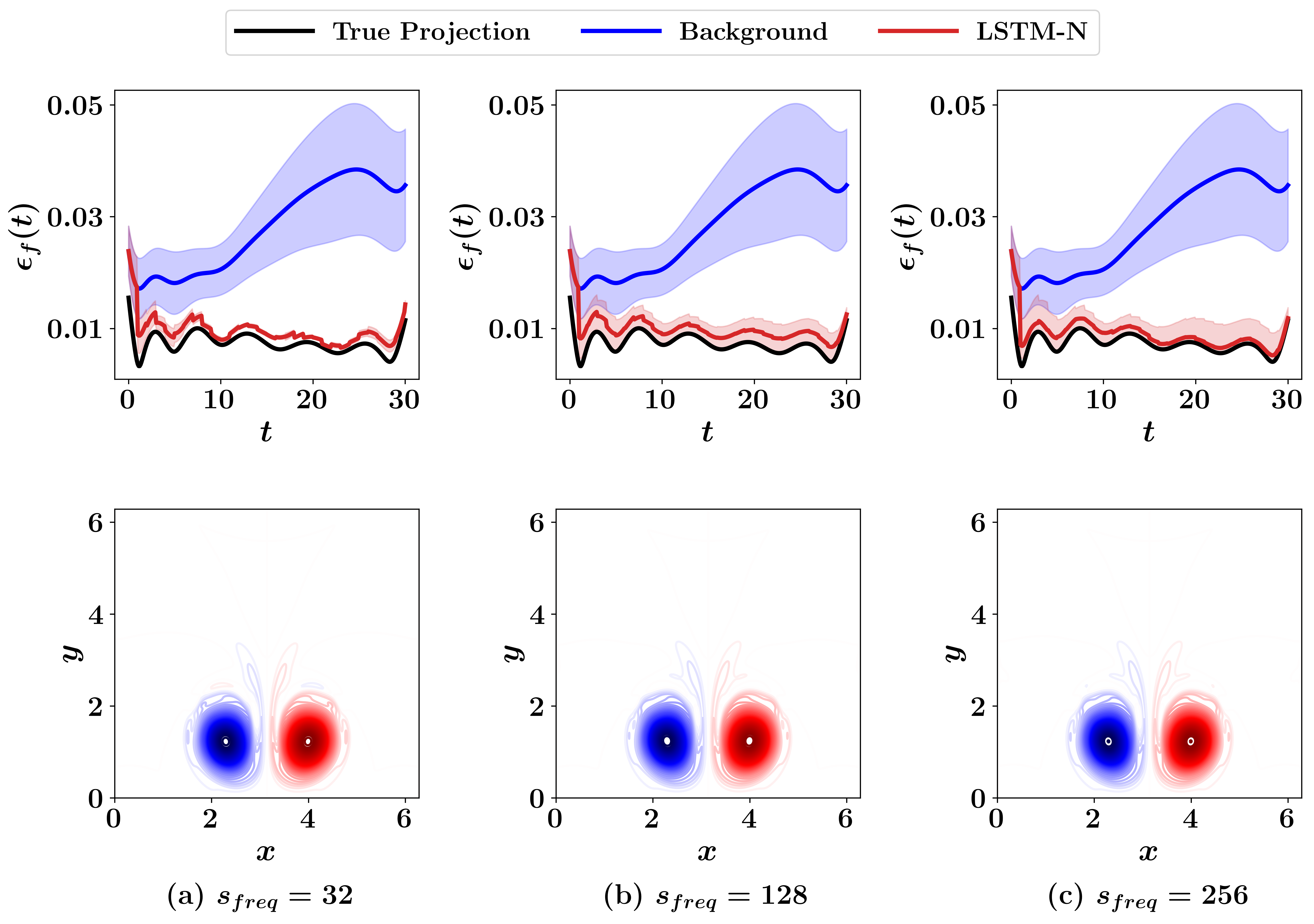}
	\caption{Comparison of resulting vorticity fields at final time as well as the line plots of root mean-squares error with time, with different number of sensors located sparsely at grid points. \textcolor{rev}{Here, $s_{freq}$ denotes the number of grid points between every two consecutive sensors.} } \label{fig:ssparse}
\end{figure}

Finally, we consider the effect of measurement sparsity on the accuracy of the presented approach. This is crucial for the trade-off between quality and quantity of sensors, since it has been shown in Section~\ref{sec:noise} that measurement noise significantly affects the LSTM-N output. For the base case, sensors are placed at every $64$ grid points. Now, we place sensors every 32 grid points, representing a denser case, as well as $128$ and $256$ grid points, representing scarcer sensors. We find that the framework is quite robust, providing very good results as illustrated in Figure~\ref{fig:ssparse}. We note here, however, that the same original LSTM cannot be utilized for testing with varying sparsity. This is because sensors sparsity controls the size of the input vector. Therefore, a new LSTM has to be re-trained for each case with the corresponding number of measurements. \textcolor{rev}{Moreover, we notice that the LSTM training suffers for the dense case with $s_{freq}=32$ ($s_{freq}$ denote the number of grid points between every two consecutive sensors). This is due to the very large input vector, requiring excessive amounts of meta-data for proper training. We also observe that some of the features in the input vector become either redundant or useless (e.g., away from the vortices). In order to reduce the size of input vector in this case (i.e., $s_{freq}=32$), we benefit from the similarities between the left and right vortices and consider measurements from only one half of the domain.} We also emphasize that compressed sensing techniques should be adopted for optimized sensors placement, rather than the simple collocated equidistant arrangement followed in the present study.

\begin{figure}[ht]	
	\centering
	\includegraphics[trim= 0 0 0 0, clip, width=1\linewidth]{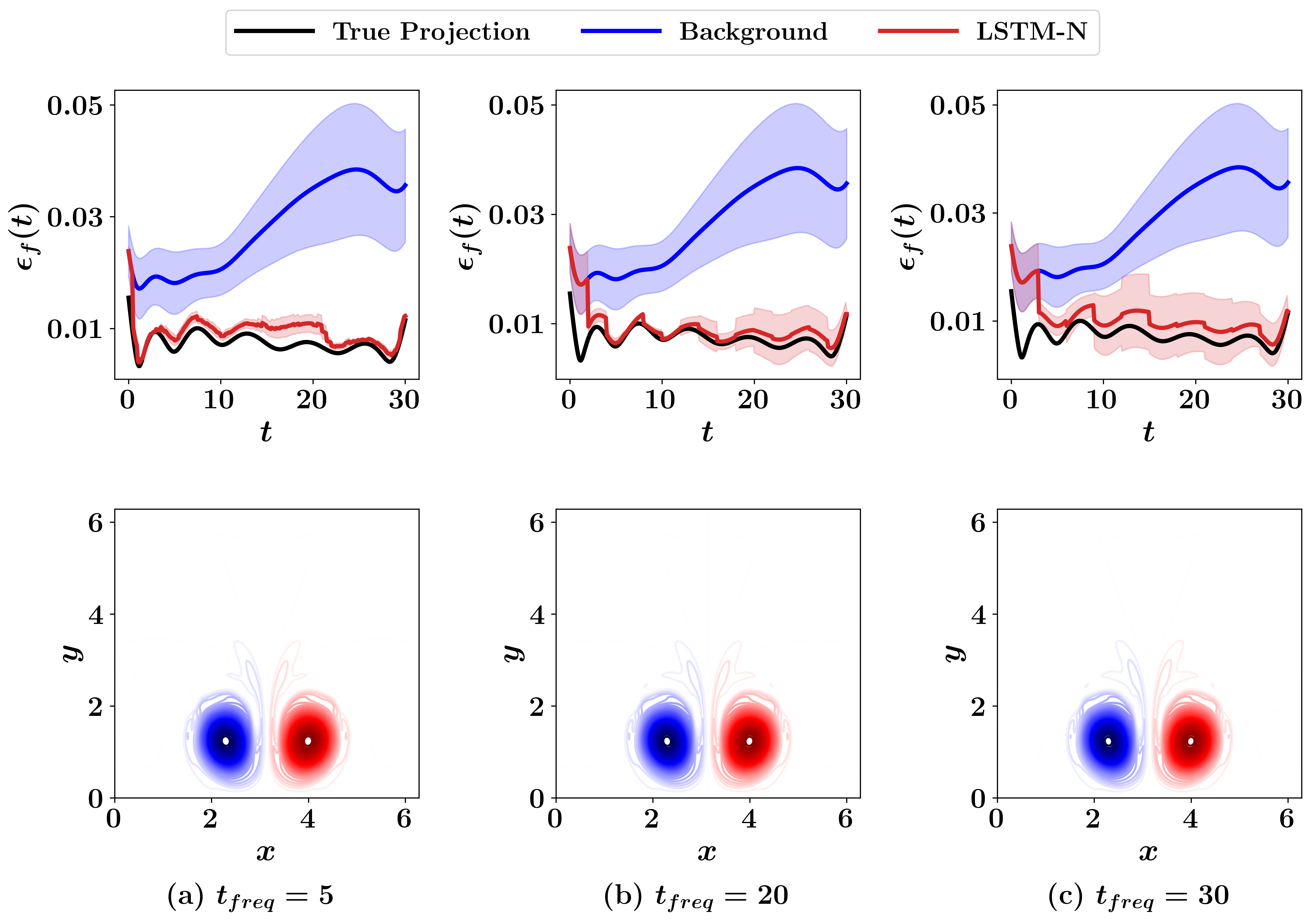}
	\caption{Comparison of resulting vorticity fields at final time as well as the line plots of root mean-squares error with time using different measurement signal frequencies. \textcolor{rev}{Here, $t_{freq}$ denotes the number of time steps between the measurement collection instants.}} \label{fig:tsparse}
\end{figure}

Regarding temporal sparsity, we collect measurement each $5$, $20$, and $30$ time-steps, compared to the reference case where measurement are collected every $10$ time-steps. We can see from Figure~\ref{fig:tsparse} that all cases yield very good predictions. \textcolor{rev}{Nevertheless, we notice that increasing $t_{freq}$ results in larger uncertainty bounds as the LSTM-N interferes at fewer time instants.} Furthermore, $\epsilon_f$ plots provide valuable insights about the capability of LSTM-N to effectively fuse measurement with background forecast to produce more accurate state estimates. For example, when measurement signals are collected every $30$ time-steps, this corresponds to $3$ time-units, meaning that the LSTM-N directly adopts the GROM prediction without correction for this amount of time, before correction is added. This is evident from Figure~\ref{fig:tsparse}c, where the red curve starts and continues with the blue curve, then a sharp reduction of the $\epsilon_f$ (and the corresponding uncertainty) is observed. This behavior is repeated as the red curve departs from the black one (corresponding to true projection) before correction is added every $\tau = 3$ time-units (i.e., $30$ time-steps). On the other hand, when more frequent measurement signals are available (e.g., every $5$ time-steps), deviation from the true projection results is less observed, as shown in Figure~\ref{fig:tsparse}a.

\section{Concluding Remarks} \label{sec:conc}
We demonstrate a machine learning based nudging approach for an idealized vortex transport problem. Such a hybrid analysis and modeling (HAM) approach is envisioned to be a promising enabler for digital twin application of an airport. Specifically, we investigate the problem of wake-vortex transport and decay as a key factor for the determination of separation distance between consecutive aircraft. Reduced order modeling based on Galerkin projection and proper orthogonal decomposition is adopted to provide computationally light models. We develop a methodology to exploit machine learning to cure model deficiency through online measurement data adopting ideas from dynamic data assimilation. Specifically, an LSTM architecture is trained to nudge prior predictions toward optimal state values using a combination of background information along with sparse and noisy observations. The proposed framework is distinguished from previous studies in the sense that it is built on the assumption that all the computing ingredients are intrinsically imperfect, including a truncated GROM model, erroneous initial conditions, and defective sensors. 

We study the effects of measurement noise and initial condition perturbation on LSTM-N behavior. The framework works sufficiently well for a wide range of noise and perturbation. Nonetheless, numerical experiments indicate relatively more dependence of performance on measurement quality (noise). Meanwhile, we find that sensors sparsity has minimal effects on results. We emphasize that the proposed framework represents a way of merging human knowledge, physics-based models, measurement information, and data-driven tools to maximize their benefits rather than discarding any of them. This becomes a key concept for building novel HAM approaches. The presented framework paves the way for viable digital twin applications to enhance airports capacities by regulating air traffic without compromising consecutive aircraft safety. Nevertheless, the scalability of the approach has yet to be tested using different vortex models and taking into account other effective factors (e.g., wind). \textcolor{rev}{For example, the wake vortex initial conditions can be related to specific aircraft types, using aircraft mass, span and speed. Also, as outlined in [8], 3D simulations are required to resolve instability mechanisms and turbulence in order to mimic wake vortex decay in a realistic way. Finally, the important effects of thermal stratification on WV descent and decay need to be considered in the transport equations.}

\section*{Acknowledgments}
This material is based upon work supported by the U.S. Department of Energy, Office of Science, Office of Advanced Scientific Computing Research under Award Number DE-SC0019290. 
O.S. gratefully acknowledges the U.S. DOE Early Career Research Program support. The work of A.R. and M.T. was supported by funding from the EU SESAR (Single European Sky ATM Research) program. 

Disclaimer: This report was prepared as an account of work sponsored by an agency of the United States Government. Neither the United States Government nor any agency thereof, nor any of their employees, makes any warranty, express or implied, or assumes any legal liability or responsibility for the accuracy, completeness, or usefulness of any information, apparatus, product, or process disclosed, or represents that its use would not infringe privately owned rights. Reference herein to any specific commercial product, process, or service by trade name, trademark, manufacturer, or otherwise does not necessarily constitute or imply its endorsement, recommendation, or favoring by the United States Government or any agency thereof. The views and opinions of authors expressed herein do not necessarily state or reflect those of the United States Government or any agency thereof.


\bibliographystyle{unsrt} 
\bibliography{references}

\end{document}